\documentclass[5p]{elsarticle}

\usepackage{amsmath, amsfonts}
\usepackage[hidelinks]{hyperref}
\usepackage{dirtytalk}
\usepackage{multirow}
\usepackage{tikz}
\usetikzlibrary{matrix}
\usetikzlibrary{positioning}
\usepackage{pgfplots}
\pgfplotsset{compat=1.15}
\usepgfplotslibrary{groupplots}
\usepackage{balance}

\renewcommand{\vec}[1]{\boldsymbol{#1}}

\newcommand{\revision}[1]{#1}

\newcommand{\throughputOverGPU}{\text{5--15}}
\newcommand{\energyOverGPU}{\text{4--13}}

\DeclareMathOperator{\DFT}{DFT}
\DeclareMathOperator{\IDFT}{IDFT}

\begin{document}

\sloppy

\title{FourierPIM: High-Throughput In-Memory Fast Fourier Transform \\ and Polynomial Multiplication}
\tnotetext[t1]{This work was supported in part by the European Research Council through the European Union's Horizon 2020 Research and Innovation Programme under Grant 757259, in part by the European Research Council through the European Union's Horizon Research and Innovation Programme under Grant 101069336, and in part by the Israel Science Foundation under Grant 1514/17.}

 \author[1]{Orian Leitersdorf\corref{cor1}}
\ead{orianl@campus.technion.ac.il}
\author[2]{Yahav Boneh}
\ead{yhb.bo@campus.technion.ac.il}
\author[3]{Gonen Gazit} 
\ead{gonn.g@campus.technion.ac.il}
\author[4]{Ronny Ronen}
\ead{ronny.ronen@technion.ac.il}
\author[5]{Shahar Kvatinsky}
\ead{shahar@ee.technion.ac.il}
\cortext[cor1]{Corresponding author}

\address[a]{Viterbi Faculty of Electrical and Computer Engineering, Technion -- Israel Institute of Technology, Haifa, 3200003, Israel}

\begin{abstract}
The Discrete Fourier Transform (DFT) is essential for various applications ranging from signal processing to convolution and polynomial multiplication. The groundbreaking Fast Fourier Transform (FFT) algorithm reduces DFT time complexity from the naive ${O(n^2)}$ to ${O(n\log n)}$, and recent works have sought further acceleration through parallel architectures such as GPUs. Unfortunately, accelerators such as GPUs cannot exploit their full computing capabilities as memory access becomes the bottleneck. Therefore, this paper accelerates the FFT algorithm using digital Processing-in-Memory (PIM) architectures that shift computation into the memory by exploiting physical devices capable of storage and logic (e.g., memristors). We propose an ${O(\log n)}$ in-memory FFT algorithm that can also be performed in parallel across multiple arrays for \emph{high-throughput batched execution}, supporting both fixed-point and floating-point numbers. Through the convolution theorem, we extend this algorithm to ${O(\log n)}$ polynomial multiplication -- a fundamental task for applications such as cryptography. We evaluate FourierPIM on a publicly-available cycle-accurate simulator that verifies both correctness and performance, and demonstrate ${\throughputOverGPU\times}$ throughput and ${\energyOverGPU\times}$ energy improvement over the NVIDIA cuFFT library on state-of-the-art GPUs for FFT and polynomial multiplication.
\end{abstract}

\maketitle

\section{Introduction}
\label{sec:introduction}

Fourier transforms are essential in a wide variety of applications such as signal processing and cryptography~\cite{F1, MATCHA}. For example, Fourier transforms can decompose an input signal (e.g., audio) according to the amplitude of the different frequencies, and can also enable fast convolution and polynomial multiplication. Particularly, the Discrete Fourier Transform (DFT) is an invertible linear transformation that operates on finite sequences of complex numbers. However, directly computing the \revision{$n$-dimensional} DFT requires $O(n^2)$ time. Fortunately, a breakthrough in the field from 1965 was the discovery of the Fast Fourier Transform (FFT)~\cite{CooleyTukey} algorithm that performs DFT with $O(n\log n)$ time by exploiting a recursive divide-and-conquer approach. Previous works have sought to further accelerate the FFT algorithm beyond $O(n \log n)$ by using GPUs for massively-parallel computation~\cite{cuFFT}.

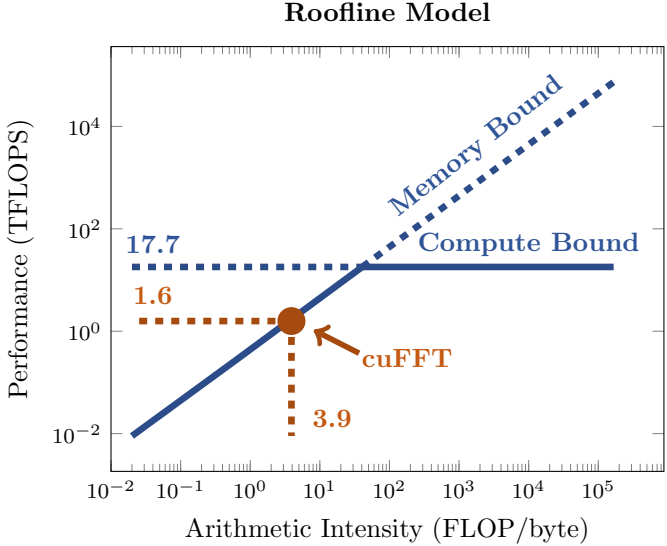
\begin{figure}[t]
    \centering
    \begin{tikzpicture}
    \begin{axis}[
        xmode=log, xmin=0.01,
        ymode=log,
        width=\linewidth,
        height=2.84in,
        xlabel = Arithmetic Intensity (FLOP/byte),
        ylabel = Performance (TFLOPS),
        title = \textbf{Roofline Model},
        every axis plot/.append style={line width=2.5pt},
        every tick label/.append style={font=\footnotesize}]
    \addplot[color={rgb:red,47;green,84;blue,151}] 
    coordinates{
    (0.02,0.009)(0.63,0.276)(40.6,18)
    };
    \addplot[color={rgb:red,47;green,84;blue,151}] 
    coordinates{
    (40.6,18)(166297.6,18)
    };
    \addplot+[color={rgb:red,197;green,90;blue,18}, only marks,mark=*,mark options={scale=2, fill={rgb:red,197;green,90;blue,18}},text mark as node=true] coordinates {
    (3.91,1.58)
    };
    \addplot[color={rgb:red,197;green,90;blue,18}, dashed] 
    coordinates{
    (3.91,1.58)(0.02,1.58)
    };
    \addplot[color={rgb:red,197;green,90;blue,18}, dashed] 
    coordinates{
    (3.91,1.58)(3.91,0.009)
    };
    \addplot[color={rgb:red,47;green,84;blue,151}, dashed] 
    coordinates{
    (40.6,18)(166297.6,73728)
    };
    \addplot[color={rgb:red,47;green,84;blue,151}, dashed] 
    coordinates{
    (0.02,18)(40.6,18)
    };
\node[rotate=38.5] at (axis cs: 1800,4000) {\textcolor[rgb]{0.184,0.329,0.592}{\textbf{Memory Bound}}};
\node[] at (axis cs: 10000,50) {\textcolor[rgb]{0.184,0.329,0.592}{\textbf{Compute Bound}}};
\node[] at (axis cs: 0.04,50) {\textcolor[rgb]{0.184,0.329,0.592}{\textbf{17.7}}};
\node[] at (axis cs: 0.04,5) {\textcolor[rgb]{0.773,0.353,0.071}{\textbf{1.6}}};
\node[] at (axis cs: 15,0.02) {\textcolor[rgb]{0.773,0.353,0.071}{\textbf{3.9}}};
\node[] at (axis cs: 180,0.3) {\textcolor[rgb]{0.773,0.353,0.071}{\textbf{cuFFT}}};
\node[] (source) at (axis cs: 50,0.3) {};
\node[] (destination) at (axis cs: 6,1.3) {};
\draw[->, line width=0.075cm, color={rgb:red,197;green,90;blue,18}](source)--(destination);
\end{axis}
\end{tikzpicture}
\caption{Roofline model for an NVIDIA RTX 3070 GPU on cuFFT; produced via the Nsight Compute profiler for batched $n=8192$ FFT.}
\label{fig:roofline}
\end{figure}

While architectures such as GPUs offer massive computational parallelism, they cannot reach full utilization for the FFT algorithm since they suffer from the memory-wall bottleneck. Figure~\ref{fig:roofline} presents a roofline model derived from the NVIDIA Nsight Compute profiling tool which accurately evaluates underlying GPU performance. The model demonstrates that the effective floating-point performance of cuFFT~\cite{cuFFT} is only $9\%$ of the theoretical maximum (1.6 TFLOPS out of 17.7 TFLOPS) since the application is memory-bound; thus, the initial $O(n)$ data transfer between the GPU cores and the GPU memory is the bottleneck. Therefore, in this work, we turn to emerging architectures that tackle the memory-wall bottleneck.

Processing-in-memory (PIM) solutions overcome the memory bottleneck by performing computation within the memory itself. We focus on digital PIM architectures~\cite{RACER, Nishil, mMPU, SIMDRAM}\footnote{\revision{PIM can also be used for highly-efficient, \emph{approximate}, matrix-vector multiplication in the \emph{analog} domain~\cite{ProcIEEEAnalog}. Previous works have demonstrated fast \emph{approximate} DFT based on this concept~\cite{AnalogFFT}, yet in this paper we focus on \emph{exact} DFT (with floating-point accuracy).}} that enable basic bitwise logic operations within the memory by exploiting underlying physical devices that inherently support both storage and computation (e.g., memristors~\cite{Memristor}). Recent works~\cite{Nishil, SIMDRAM, AritPIM, SIMPLER, Ameer} have demonstrated in-memory vectored arithmetic with massive throughput, with AritPIM~\cite{AritPIM} proposing a suite of arithmetic operations for both fixed-point and floating-point numbers. While previous works have attempted similar transformations using PIM~\cite{CryptoPIM, MeNTT, Crafft}, they require complex custom \emph{near-memory} periphery that does not support parallel batched execution and thus limits the overall throughput to \emph{less} than existing GPU solutions.

In this paper, we exploit the massive arithmetic throughput of digital PIM towards \emph{high-throughput} FFT that is performed entirely through \emph{in-memory} operations. We design an in-memory in-place FFT algorithm that exploits the full parallelism within a single memory array, and attains $O(\log n)$ time complexity for an $n$-dimensional FFT. Furthermore, this algorithm may be executed in parallel across all of the memory arrays, thereby providing high throughput for batched FFT execution. We extend the algorithm to in-array polynomial multiplication by utilizing the convolution theorem, attaining $O(\log n)$ time. The proposed algorithms are evaluated on a \emph{publicly-available cycle-accurate simulator} that both verifies correctness and also measures performance (based on the parameters from the RACER~\cite{RACER} architecture), and are compared to cuFFT~\cite{cuFFT} on state-of-the-art GPUs (NVIDIA RTX 3070 and NVIDIA A100). This paper contributes:
\begin{itemize}
    \item \emph{Complex Arithmetic:} Extends AritPIM~\cite{AritPIM} to vectored in-memory arithmetic for \emph{complex numbers}.
    \item \emph{In-Memory FFT:} Proposes the first \emph{high-throughput} FFT algorithm for PIM that is based completely on in-memory operations, attaining $O(\log n)$ time while also enabling parallel batched execution.
    \item \emph{Convolution and Polynomial Multiplication:} Extends the proposed FFT algorithms to convolution and polynomial multiplication while also exploiting additional optimizations.
    \item \emph{Evaluation:} Demonstrates a $\throughputOverGPU\times$ throughput and $\energyOverGPU\times$ energy improvement over cuFFT~\cite{cuFFT}. 
\end{itemize}

This paper is organized as follows. Background is provided in Section~\ref{sec:PIM} on digital PIM techniques and in Section~\ref{sec:fourier} on Fourier transforms. Section~\ref{sec:proposed} proposes the in-memory FFT algorithm, and Section~\ref{sec:extensions} extends this towards convolution and polynomial multiplication through the convolution theorem. Section~\ref{sec:evaluation} evaluates the proposed algorithms and Section~\ref{sec:conclusion} concludes this paper. 

\section{Digital Processing-in-Memory (PIM)}
\label{sec:PIM}

Processing-in-memory (PIM) architectures are rapidly emerging to combat the memory bottleneck by integrating computation within the memory itself. PIM architectures enable the computer to perform vectored arithmetic operations within the memory without transferring the data to dedicated compute units (e.g., CPU), thereby nearly eliminating the data transfer between compute and memory. PIM architectures are generally based on physical devices that inherently support both information storage and basic logic operations, providing massive parallelism for basic bitwise operations which is then expanded towards high-throughput vectored in-memory arithmetic.

\begin{figure}
    \centering
    \includegraphics[width=\linewidth]{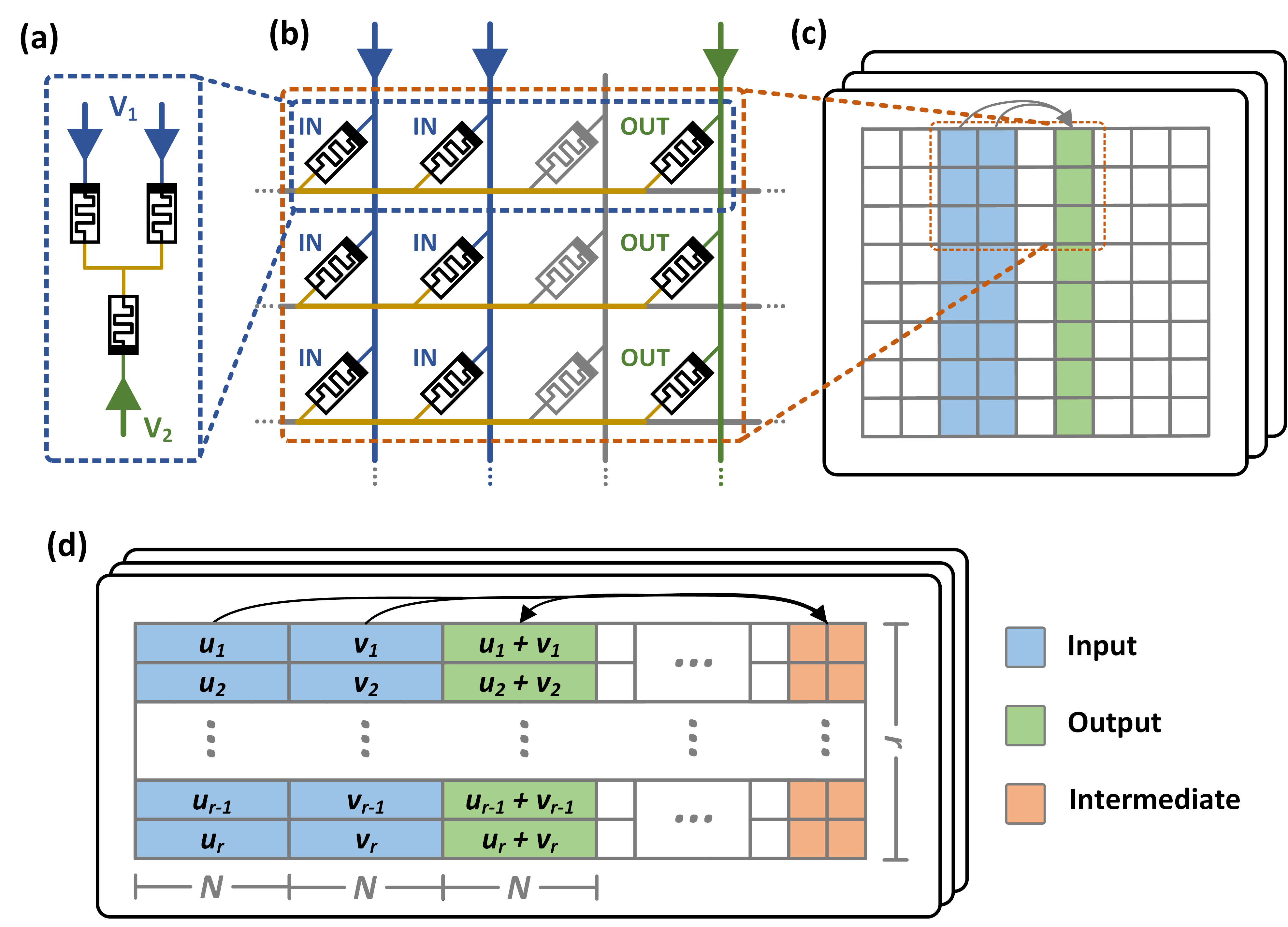}
    \caption{(a) Stateful logic performed between input (top) and output (bottom) memristors. (b) Stateful logic within rows of a crossbar by applying voltages on bitlines. (c) Abstract computational model that provides bitwise operations on columns with ${O(1)}$ time. (d) Utilization of bitwise column operations towards vectored arithmetic.}
    \label{fig:PIM}
\end{figure}

\subsection{Digital Memristive PIM}
\label{sec:PIM:memristive}

Digital memristive PIM architectures~\cite{RACER, Nishil, mMPU} exploit memristors~\cite{Memristor}: two-terminal physical devices (similar to resistors) that possess an internal resistance which may be modified with an electric current. Memristors may store binary information through their resistance (usually high resistance for logical zero and low resistance for logical one), writing data with an applied current and reading data by applying a low voltage and measuring the current. Memristors also support \emph{digital logic} in the resistance domain (stateful logic~\cite{MemristiveLogic}) using the memristors themselves as the building blocks~\cite{IMPLY, MAGIC, FELIX}. For memristors arranged in a circuit as illustrated in Figure~\ref{fig:PIM}(a), the state of the bottom memristor will change conditional on the states of the top two memristors (e.g., NOR~\cite{MAGIC}). The same circuit exists within a single row of a memristive crossbar array (vertical bitlines, horizontal wordlines, and memristors at all junctions), as shown in Figure~\ref{fig:PIM}(b). Therefore, by applying voltages on bitlines, up to all of the rows simultaneously perform a logic gate between the selected memristors. An abstraction of this model is that a crossbar array is a binary matrix of memory that supports bitwise logic on arbitrary columns with $O(1)$ time complexity (e.g., bitwise NOR of two columns stored in a third column is performed in a single clock cycle), see Figure~\ref{fig:PIM}(c). Furthermore, by applying voltages on wordlines rather than bitlines, logic operations can be performed on rows of a crossbar array. The memory consists of many crossbar arrays, each typically of size $1024 \times 1024$~\cite{Bitlet}. 

\subsection{Element-Parallel Arithmetic}
\label{sec:PIM:arithmetic}

Element-parallel arithmetic extends the bitwise parallelism provided by digital PIM towards vectored arithmetic~\cite{AritPIM, SIMPLER, Nishil, Ameer}. For example, the addition of vectors $\vec{u} = (u_1, \hdots, u_r)$ and $\vec{v} = (v_1, \hdots, v_r)$, each $N$-bit numbers (fixed or floating), starts with the vectors arranged in an $r \times c$ crossbar array where the $i^{th}$ row contains $u_i$ and $v_i$, see Figure~\ref{fig:PIM}(d). The element-parallel arithmetic computes the vector addition $\vec{u} + \vec{v}$ by computing $u_i + v_i$ independently in each row and in parallel across all rows. This is performed via \emph{bit-serial} computation (also known as \say{single-row}~\cite{SIMPLER}): $N$-bit addition is converted to a serial sequence of basic logic gates (e.g., NOR), and each logic gate is serially performed within each row, yet in parallel across all rows. Additional columns may store intermediate results utilized for the arithmetic (e.g., the carry bit in addition).\footnote{For simplicity, we do not discuss intermediate arithmetic cells in the text, yet they are evaluated in the simulations.} AritPIM~\cite{AritPIM} recently proposed a suite of arithmetic operations for both fixed-point and floating-point numbers; we adopt these algorithms.

\section{Fourier Transforms}
\label{sec:fourier}

Fourier transforms are essential for numerous tasks ranging from signal processing to convolution~\cite{ F1, MATCHA, nussbaumer1981fast}. The Discrete Fourier Transform (DFT)~\cite{IntroToAlgorithms} operates on sequences of complex numbers, producing alternative representations in the frequency domain. The Fast Fourier Transform (FFT)~\cite{CooleyTukey, IntroToAlgorithms} performs DFT at a significantly lower time complexity: $O(n\log n)$ instead of $O(n^2)$.

\subsection{Discrete Fourier Transform (DFT)}
\label{sec:fourier:DFT}

The Discrete Fourier Transform (DFT) is an invertible linear transformation that converts an \revision{$n$-dimensional} input sequence $x_0, \hdots, x_{n-1} \in \mathbb{C}$ to $X_0, \hdots, X_{n-1}= \DFT(x_0, \hdots, x_{n-1})$ according to
\begin{equation}
X_k = \sum_{j=0}^{n-1} x_j \cdot {\omega_n}^{j\cdot k},
\label{eq:DFT}
\end{equation}
where $\omega_n = e^{-2\pi i/n}$ and $i=\sqrt{-1}$. The inverse DFT reverses this to compute $x_0, \hdots, x_{n-1}=\IDFT(X_0, \hdots, X_{n-1})$ by
\begin{equation}
x_j = \frac{1}{n} \cdot \sum_{k=0}^{n-1} X_k \cdot {\omega_n}^{-j\cdot k}.
\label{eq:IDFT}
\end{equation}
While directly computing the DFT and inverse DFT requires $O(n^2)$ operations ($O(n)$ operations for each of the $O(n)$ elements), the Fast Fourier Transform (FFT) reduces this to $O(n\log n)$ operations overall.

\subsection{Fast Fourier Transform (FFT)}
\label{sec:fourier:FFT}

\begin{figure}
    \centering
    \includegraphics[width=\linewidth, trim={0cm, 0.5cm, 0cm, 0cm}]{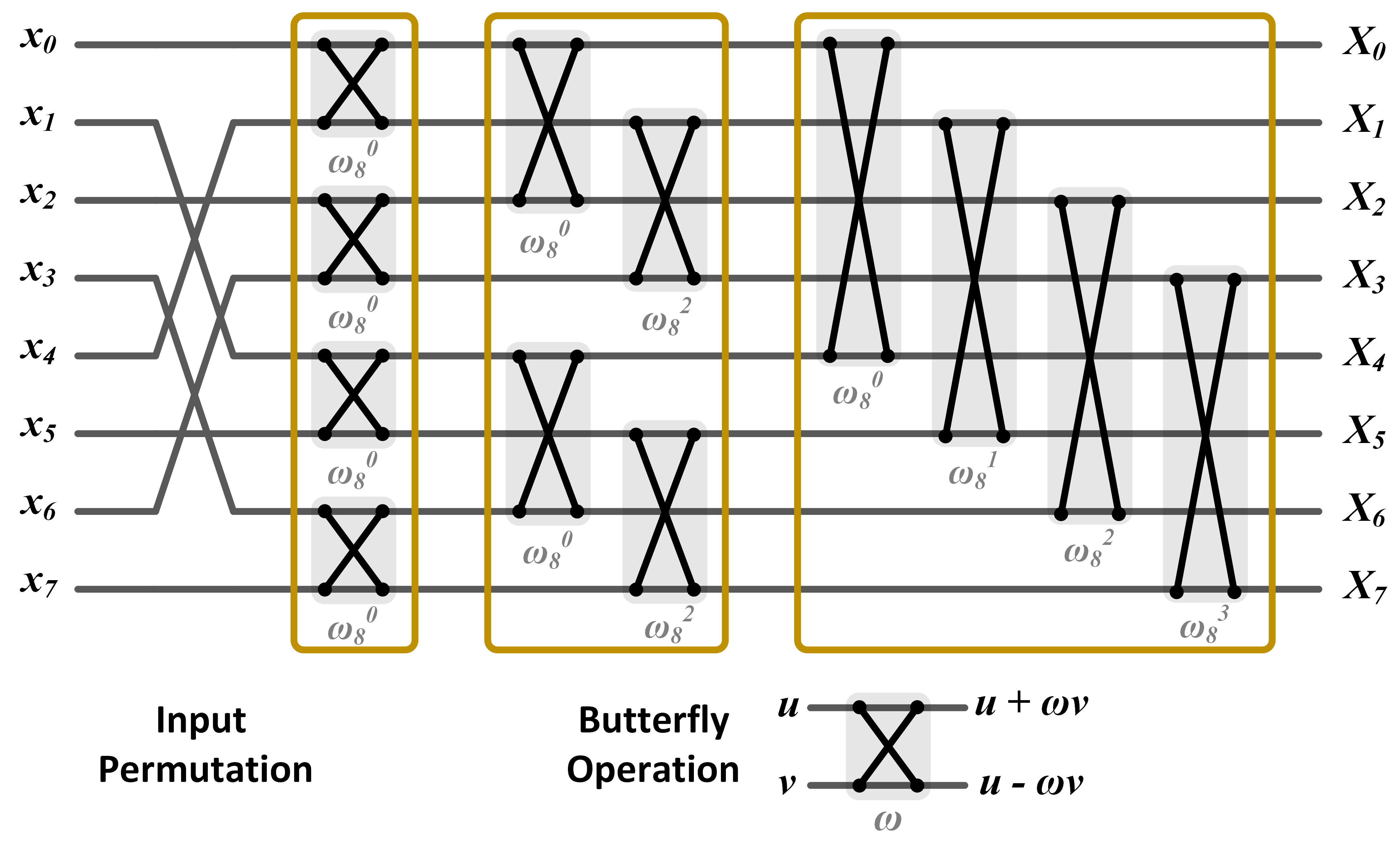}
    \caption{Parallel representation of the FFT algorithm for ${n=8}$~\cite{IntroToAlgorithms}. Each butterfly operation operates on two elements from the sequence at a time, and every group (yellow) of ${n/2}$ operations may be performed simultaneously.}
    \label{fig:fft}
\end{figure}

The FFT is based upon a recursive divide-and-conquer approach that separates the input by even and odd indices,
\begin{multline}
X_k = \sum_{j=0}^{n/2-1}x_{2j}\cdot {\omega_n}^{2j\cdot k} + \sum_{j=0}^{n/2-1}x_{2j + 1}\cdot {\omega_n}^{(2j + 1)\cdot k} \\
= \underbrace{\sum_{j=0}^{n/2-1}x_{2j}\cdot {\omega_{n/2}}^{j\cdot k}}_{\DFT(x_0, x_2, \hdots)_{k\text{ mod } n/2}} + {\omega_n}^{k}\cdot\underbrace{\sum_{j=0}^{n/2-1}x_{2j + 1}\cdot {\omega_{n/2}}^{j\cdot k}}_{\DFT(x_1, x_3, \hdots)_{k\text{ mod } n/2}}.
\label{eq:FFT}
\end{multline}
Therefore, we find that $\DFT(x_0, \hdots, x_{n-1})$ can be derived from $\DFT(x_0, x_2, \hdots, x_{n-2})$, $\DFT(x_1, x_3, \hdots, x_{n-1})$ and $O(n)$ additional additions and multiplications, recursively leading to $O(n\log n)$ time:
\revision{
\begin{flalign}
&T(n) = 2 \cdot T\left(n/2\right) + O(n), \\
&\implies T(n) = O(n) + \underbrace{2 \cdot O\left(\frac{n}{2}\right)}_{O(n)} + \underbrace{4 \cdot O\left(\frac{n}{4}\right)}_{O(n)} + \cdots, \\
&\implies T(n) = O(n \log n),
\end{flalign}
as there are $\log(n)$ elements.} An equivalent FFT algorithm is shown in Figure~\ref{fig:fft} by recursively computing the FFT for smaller sequences and combining the results with $O(n)$ \emph{butterfly operations}~\cite{IntroToAlgorithms}. This approach begins with a permutation of the input~\cite{IntroToAlgorithms} and continues with $\log n$ groups of $n/2$ butterfly operations (each operating on two elements of the sequence). As each group contains independent operations, this enables computing the FFT algorithm with $O(\log n)$ time on a parallel processor. 

\section{Proposed In-Memory FFT}
\label{sec:proposed}

In this section, we map the parallel FFT algorithm to a memristive crossbar array. By fully exploiting the parallelism provided by memristive logic, we attain $O(\log n)$ time complexity for batched execution \revision{as each group of butterfly operations is performed with $O(1)$ time complexity and there are $\log n$ groups}. We focus on \emph{in-place} FFT (output overrides the input); if input preservation is desired, then the inputs can be copied to different columns before starting. We explore three configurations for storing the input sequence within an $r \times c$ crossbar:

\begin{itemize}
    \item \emph{$r$-configuration ($n=r$):} $x_0, \hdots, x_{n-1}$ are stored over $N$ columns with one element in each row.
    \item \emph{$2r$-configuration ($n=2r$):} $x_0, \hdots, x_{n-1}$ are stored over $2N$ columns, two elements per row (organized according to a snake pattern).
    \item \emph{$2r\beta$-configuration ($n=2r\beta$) for $\beta>0$ constant:} The input sequence $x_0, \hdots, x_{n-1}$ is stored over $2N\beta$ columns with $2\beta$ elements per row (snake pattern).
\end{itemize}
This section is organized as follows. Section~\ref{sec:proposed:complex} extends AritPIM~\cite{AritPIM} to complex arithmetic and Section~\ref{sec:proposed:butterfly} proposes an element-parallel butterfly operation. Lastly, Sections~\ref{sec:proposed:r},~\ref{sec:proposed:2r}, and \ref{sec:proposed:2rbeta} gradually develop the proposed algorithms for $r$, $2r$, and $2r\beta$ (respectively).

\subsection{Complex Arithmetic}
\label{sec:proposed:complex}

We choose to represent complex numbers according to their rectangular form ($a + bi$) rather than the polar form ($r$ and $\theta$) as this simplifies addition. We store $N$-bit complex floating-point numbers in the memory as the concatenation of two $N/2$-bit real floating-point numbers representing $a$ and $b$. Complex addition/subtraction is derived from real operations by
\begin{equation}
(a + bi) \pm (a' + b'i) = (a \pm a') + (b \pm b')i.
\end{equation}
Similarly, multiplication is performed according to
\begin{equation}
(a + bi) \cdot (a' + b'i) = (aa' - bb') + (ab' + a'b)i.
\end{equation}

\subsection{Butterfly Operation}
\label{sec:proposed:butterfly}

We extend the complex arithmetic towards an element-parallel butterfly operation: receives vectors $\vec{u}, \vec{v}, \vec{\omega}$ stored with a single element in each row, and outputs $\vec{u} + \vec{\omega} \odot \vec{v}$ and $\vec{u} - \vec{\omega} \odot \vec{v}$, where $\odot$ represents element-wise multiplication. This is performed by multiplying $\vec{\omega}$ and $\vec{v}$, adding the result to $\vec{u}$ to get $\vec{u} + \vec{\omega} \odot \vec{v}$, and subtracting the result from $\vec{u}$ to get $\vec{u} - \vec{\omega} \odot \vec{v}$. We also support in-place computation that overrides $\vec{u},\vec{v}$ with $\vec{u} + \vec{\omega} \odot \vec{v},\vec{u} - \vec{\omega} \odot \vec{v}$. \revision{Notice that since vectored arithmetic operations are supported with $O(1)$ time complexity~\cite{AritPIM}, then we find that the vectored butterfly operation is also $O(1)$ time as it involves $O(1)$ arithmetic operations.}

Thus far, we have developed a high-throughput vectored butterfly operation: all of the rows may perform butterfly operations in parallel -- each row representing a single operation. We continue in this section by reducing the FFT algorithm to vectored butterfly operations that are performed within the crossbar, thereby exploiting the high throughput towards FFT acceleration.

\subsection{\texorpdfstring{$r$}{r}--FFT}
\label{sec:proposed:r}

\begin{figure*}
    \centering
    \includegraphics[width=0.925\linewidth, trim={0cm, 0.3cm, 0cm, 0cm}]{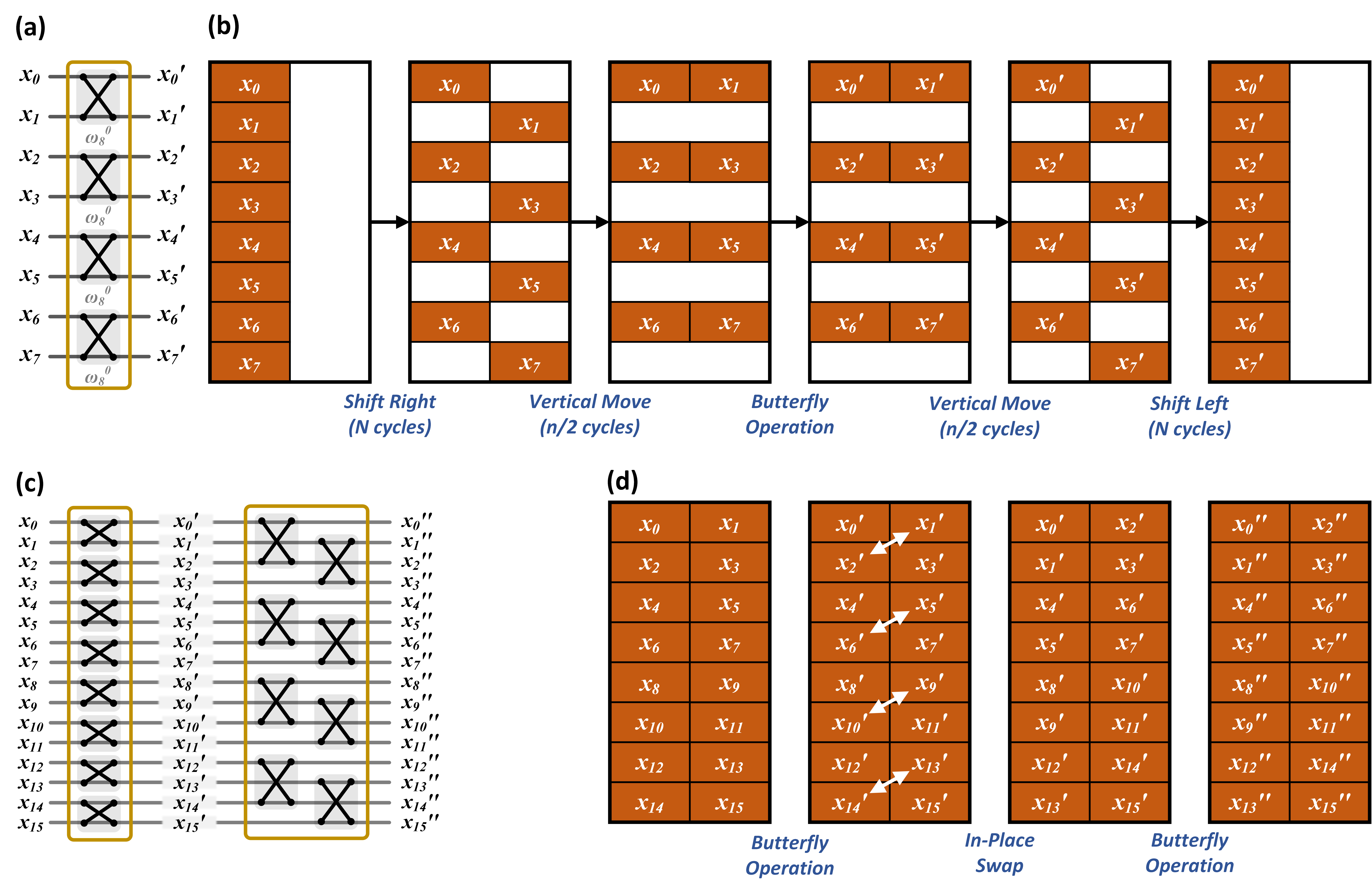}
    \caption{(a, b): Example ${r}$--FFT steps for the first group when ${n=8}$. (c, d) Example ${2r}$--FFT transition between the first and second groups for ${n=16}$; this enables double the dimension with identical area and similar latency. For clarity, ${x_0, x_1, \hdots, x_{n-1}}$ shown in the illustration are after input permutation.}
    \label{fig:rFFT}
    \vspace{-10pt}
\end{figure*}

We map the algorithm from Figure~\ref{fig:fft} to an $r$-configuration crossbar, while utilizing element-parallel butterfly operations. 

The invariant in the algorithm is storing the sequence $x_0, \hdots, x_{n-1}$ in $r$-configuration between the groups of Figure~\ref{fig:fft}. We start by permuting the sequence according to the input permutation step using logic operations on rows of the crossbar array (e.g., swap is performed using NOT gates). Then, for each yellow group from Figure~\ref{fig:fft}, we perform the following steps (example shown in Figure~\ref{fig:rFFT}):
\begin{enumerate}
    \item Arrange the sequence such that the inputs of each butterfly are in the same row by shifting half of the sequence to the right and upwards (NOT gates).
    \item Perform the butterfly operation in parallel across the $r/2$ rows that contain inputs. The \emph{constants} ${\omega_n}^j$ are produced using standard write operations\footnote{These can be performed in parallel across all crossbars as they are for constants, and they also have negligible latency overall.}.
    \item Move the sequence back into $r$-configuration by shifting downwards and then to the left.
\end{enumerate}
Therefore, we find that all of the groups are performed using the parallelism provided from the element-parallel butterfly operations, and that the entire FFT is performed within the crossbar. Yet, there remain two aspects of this solution that can be improved: (1) the area footprint is double the size of the sequence itself, (2) the butterfly operation only operates on half of the rows in parallel. The proposed $2r$-configuration algorithm exploits both of these aspects to provide $n=2r$ dimensional FFT with nearly identical area and latency. 

\subsection{\texorpdfstring{$2r$}{2r}--FFT}
\label{sec:proposed:2r}

The $2r$-configuration FFT improves the $r$-configuration FFT by avoiding the intermediate representation in transitions between groups. Consider the parallel FFT algorithm for $n=16$ (shown in Figure~\ref{fig:rFFT}(c)) instead of $n=8$. Rather than start with a single element per row, we start with two elements stored per row (snake format) -- this already corresponds to the format for the first butterfly operation. To transition between the formats for consecutive butterfly operations, an in-place swap is performed between pairs of elements (see Figure~\ref{fig:rFFT}(d)).

\subsection{\texorpdfstring{$2r\beta$}{2rbeta}--FFT}
\label{sec:proposed:2rbeta}

In this configuration, we aim to maximize the supported FFT dimension by storing the input sequence over multiple pairs of $N$ columns. Specifically, for $n=2r\beta$, we store the input in a snake pattern across $2 N\beta$ columns. Each group of $2N$ columns constitutes a unit that performs butterfly operations, and the in-place swaps are performed serially amongst the different units. We also utilize swaps between different units in the form of column operations. While in the standard $2r\beta$ model the butterfly operations are performed serially amongst the units, we also evaluate an extension of $2r\beta$ that utilizes partitions~\cite{PartitionPIM} to enable parallel execution.

\section{Proposed Polynomial Multiplication}
\label{sec:extensions}

This section expands the in-memory FFT algorithms towards 1D convolution and polynomial multiplication \revision{with $O(\log n)$ time}. Without loss of generality, we discuss only polynomial multiplication as 1D convolution is equivalent. We continue by providing background on the convolution theorem and then detailing our optimized in-memory polynomial multiplication algorithm.

The convolution theorem describes a fundamental relationship between polynomial multiplication and Fourier transforms~\cite{IntroToAlgorithms}. A polynomial $A$ of degree $n-1$ is defined by $A(x) = \sum_{j=0}^{n-1}a_ix^i$, and the product of polynomials $A$ and $B$ is the polynomial $C$ that satisfies $C(x) = A(x) \cdot B(x)$ for all $x$. Polynomial multiplication is traditionally performed as a convolution of the coefficients (e.g., $(2 + 3x) * (1 + 4x) = 2 + (8 + 3)x + 12x^2$) in $O(n^2)$ time. Yet, if polynomials are represented according to their values at $n$ fixed points rather than the coefficients (e.g., storing $A(0)$ and $A(1)$ instead of $a_0$ and $a_1$), then polynomial multiplication is merely multiplying the values of the polynomials at the fixed points in $O(n)$ time. Furthermore, by choosing the fixed points as the $n$ roots of unity ${\omega_n}^0,\hdots,{\omega_n}^{n-1}$, converting a coefficient representation to and from a point representation is precisely a DFT and inverse DFT (notice that Eq.~(\ref{eq:DFT}) is evaluating the polynomial at ${\omega_n}^k$) and thus can be performed with $O(n\log n)$ time. Overall,
\begin{multline}
(c_0, \hdots, c_{n-1}) = \\ \IDFT(\DFT(a_0, \hdots, a_{n-1}) \odot \DFT(b_0, \hdots, b_{n-1})),
\end{multline}
where $\odot$ represents element-wise multiplication.\footnote{Note that this produces $C$ of degree at most $n$. If degree up to $2n$ is desired, $A$ and $B$ may be padded with an additional $n$ zeros.}

We utilize the convolution theorem alongside the proposed FFT algorithm towards fast in-memory polynomial multiplication. We assume that the input polynomials are provided in $r$, $2r$, or $2r\beta$ configuration each, and perform the following steps: (1) compute FFT for the coefficients of each polynomial, (2) compute the element-wise product of the FFT outputs, and (3) compute an inverse FFT on the product (inverse FFT algorithm is similar to FFT). Note that as we compute both the FFT and the inverse FFT, then there is no need to perform the initial input permutations as they cancel out~\cite{IntroToAlgorithms}. \revision{Since the proposed FFT possesses $O(\log n)$ time complexity, then this polynomial multiplication also has $O(\log n)$ time since it involves a constant number of FFTs and the element-wise product can be performed with $O(1)$ time using AritPIM~\cite{AritPIM}.}

We further accelerate polynomial multiplication for real coefficients by performing the two FFTs in step (1) together by combining two $n$-dimensional FFTs with real coefficients into a single $n$-dimensional complex FFT. Specifically, for real input sequences $x_1, \hdots, x_n \in \mathbb{R}$ and $y_1, \hdots y_n \in \mathbb{R}$, we define complex sequence $z_1, \hdots, z_n \in \mathbb{C}$ according to $z_k = x_k + i \cdot y_k$. The FFTs of the initial sequences are derived according to,
\begin{equation}
X_k = \frac{1}{2} \cdot \left[\overline{Z_{n-k}}+Z_k\right], \quad\quad Y_k = \frac{i}{2} \cdot \left[\overline{Z_{n-k}}-Z_k\right].
\end{equation}
With regards to the in-memory implementation: (1) complex conjugate is performed by flipping the sign bit of the imaginary part, (2) multiplication by $i$ is performed by simulating a swap of the real and imaginary parts, followed by an inversion of the sign of the imaginary part, (3) division by 2 is performed by decrementing the exponent, and (4) the values of $Z_{n-k}$ are derived by reversing the order of $Z$ using swap operations.

\section{Evaluation}
\label{sec:evaluation}

\begin{table}[t]
    \centering
    \caption{Evaluation Parameters}
    \vspace{5pt}
    \begin{tabular}{|c|l|}
        \hline
        \textbf{Configuration} & \textbf{Parameters} \\
        \hline
        \hline
        \multirow{4}{*}{FourierPIM--8/40} &
        \emph{Memory Size:} 8/40 GB \\
        & \emph{Crossbar Size:} $1024 \times 1024$~\cite{Bitlet} \\
        & \emph{Clock Frequency:} 333.3 MHz~\cite{RACER} \\
        (with partitions) & \emph{Gate Energy (fJ):} 6.4 fJ~\cite{RACER} \\
        & (\emph{Partition Count:} up to 4.) \\
        \hline
        \multirow{3}{*}{cuFFT (RTX 3070)} & \emph{Number of Cores:} 5888\\
        & \emph{Memory Size:} 8 GB \\
        & \emph{Memory Bandwidth:} 448 GB/s \\
        \hline
        \multirow{3}{*}{cuFFT (A100)} & \emph{Number of Cores:} 6912\\
        & \emph{Memory Size:} 40 GB \\
        & \emph{Memory Bandwidth:} 1,555 GB/s \\
        \hline
    \end{tabular}
    \label{tab:params}
\end{table}

\begin{figure}[t]
    \centering
  \begin{tikzpicture}
    \begin{groupplot}[
      group style={group size=2 by 2, horizontal sep=1.6cm,
      vertical sep=1.3cm},
      width=4.5cm, height=4cm
    ]
    \nextgroupplot[
        xmode=log,
        ymode=log,
        log basis x={2},
        xtick={2048, 4096, 8192},
        xlabel={\footnotesize $n$},
        ylabel={\footnotesize Tput. (FFT/sec)},
        every axis plot/.append style={ultra thick},
        every tick label/.append style={font=\footnotesize}
    ]
    \addplot[color=black]
    coordinates {(2048, 12318669.2)(4096, 6159668.5)(8192, 3080107.4)};\label{plots:RTX3070};
    \addplot[color=brown]
    coordinates {(2048, 42192811.0)(4096, 21162910.4)(8192, 10496051.8)};\label{plots:A100};
    \addplot[color=blue]
    coordinates {(2048, 3.65E+07)(4096, 1.70E+07)(8192, 7.90E+06)};\label{plots:PolyPIM8};
    \addplot[color=blue, dashed]
    coordinates {(2048, 3.65E+07)(4096, 2.90E+07)};\label{plots:PolyPIM8P};
    \addplot[color=orange]
    coordinates {(2048, 1.82E+08)(4096, 8.48E+07)(8192, 3.95E+07)};\label{plots:PolyPIM40};
    \addplot[color=orange, dashed]
    coordinates {(2048, 1.82E+08)(4096, 1.45E+08)};\label{plots:PolyPIM40P};
    
    \coordinate (top) at (rel axis cs:0,1);
    
    \nextgroupplot[
        xmode=log,
        ymode=log,
        log basis x={2},
        xtick={2048, 4096, 8192},
        xlabel={\footnotesize $n$},
        ylabel={\footnotesize Tput. / Watt},
        every axis plot/.append style={ultra thick},
        every tick label/.append style={font=\footnotesize}
    ]
    \addplot[color=black]
    coordinates {(2048, 74700.3)(4096, 37118.1)(8192, 18827.4)};
    \addplot[color=brown]
    coordinates {(2048, 174622.2)(4096, 86318.1)(8192, 42210.3)};
    \addplot[color=blue]
    coordinates {(2048, 6.28E+05)(4096, 2.89E+05)(8192, 1.33E+05)};
    \addplot[color=blue, dashed]
    coordinates {(2048, 6.28E+05)(4096, 2.89E+05)(8192, 1.33E+05)};
    \addplot[color=orange, dashed]
    coordinates {(2048, 6.28E+05)(4096, 2.89E+05)(8192, 1.33E+05)};
    \addplot[color=orange, dashed]
    coordinates {(2048, 6.28E+05)(4096, 2.89E+05)(8192, 1.33E+05)};

    \nextgroupplot[
        xmode=log,
        ymode=log,
        log basis x={2},
        xtick={2048, 4096, 8192, 16384},
        xlabel={\footnotesize $n$},
        ylabel={\footnotesize Tput. (FFT/sec)},
        every axis plot/.append style={ultra thick},
        every tick label/.append style={font=\footnotesize}
    ]
    \addplot[color=black]
    coordinates {(2048, 24773607.3)(4096, 12388881.3)(8192, 6162492.6)(16384, 1519123.2)};
    \addplot[color=brown]
    coordinates {(2048, 83313199.3)(4096, 41711158.6)(8192, 20718861.6)(16384, 5271570.2)};
    \addplot[color=blue]
    coordinates {(2048, 7.89E+07)(4096, 3.72E+07)(8192, 1.76E+07)(16384, 8.29E+06)};
    \addplot[color=blue, dashed]
    coordinates {(2048, 7.89E+07)(4096, 5.55E+07)(8192, 3.55E+07)};
    \addplot[color=orange]
    coordinates {(2048, 3.94E+08)(4096, 1.86E+08)(8192, 8.78E+07)(16384, 4.15E+07)};
    \addplot[color=orange, dashed]
    coordinates {(2048, 3.94E+08)(4096, 2.77E+08)(8192, 1.78E+08)};
    
    \coordinate (top2) at (rel axis cs:0,1);
    
    \nextgroupplot[
        xmode=log,
        ymode=log,
        log basis x={2},
        xtick={2048, 4096, 8192, 16384},
        xlabel={\footnotesize $n$},
        ylabel={\footnotesize Tput. / Watt},
        every axis plot/.append style={ultra thick},
        every tick label/.append style={font=\footnotesize}
    ]
    \addplot[color=black]
    coordinates {(2048, 140802.1)(4096, 72901.7)(8192, 36267.6)(16384, 7700.3)};
    \addplot[color=brown]
    coordinates {(2048, 297254.7)(4096, 163543.8)(8192, 77033.3)(16384, 17620.0)};
    \addplot[color=blue]
    coordinates {(2048, 1.90E+06)(4096, 8.73E+05)(8192, 4.04E+05)(16384, 1.88E+05)};
    \addplot[color=blue, dashed]
    coordinates {(2048, 1.90E+06)(4096, 8.73E+05)(8192, 4.04E+05)(16384, 1.88E+05)};
    \addplot[color=orange, dashed]
    coordinates {(2048, 1.90E+06)(4096, 8.73E+05)(8192, 4.04E+05)(16384, 1.88E+05)};
    \addplot[color=orange, dashed]
    coordinates {(2048, 1.90E+06)(4096, 8.73E+05)(8192, 4.04E+05)(16384, 1.88E+05)};

    \coordinate (bot) at (rel axis cs:1,0);
    
    \end{groupplot}
    
    \path (top|-current bounding box.north)--
          coordinate(legendpos)
          (bot|-current bounding box.north);
    \matrix[
        matrix of nodes,
        anchor=south,
        draw,
        inner sep=0.1em,
        draw
      ] at([xshift=-3ex,yshift=5ex]legendpos)
      {
    \ref{plots:PolyPIM8}& FourierPIM--8 &[5pt]
    \ref{plots:PolyPIM8P}& FourierPIM--8 w/ partitions&[5pt] \\
    \ref{plots:PolyPIM40}& FourierPIM--40 &[5pt]
    \ref{plots:PolyPIM40P}& FourierPIM--40  w/ partitions&[5pt] \\
    \ref{plots:RTX3070}& RTX 3070 &[5pt]
    \ref{plots:A100}& A100 &[5pt]\\};
    \node[] at ([yshift=2ex]legendpos) {Full Precision (64-bit floating)};
    \node[] at ([yshift=2ex, xshift=24ex]top2) {Half Precision (32-bit floating)};
    \node[] at ([yshift=2ex,xshift=-6ex]top) {(a)};
    \node[] at ([yshift=2ex,xshift=-6ex]top2) {(b)};
  \end{tikzpicture}
  \vspace{-10pt}
  \caption{Comparison of the proposed FFT algorithms to the baseline cuFFT for (a) full precision and (b) half precision.}
  \label{fig:results:fft}
  \vspace{-10pt}
\end{figure}
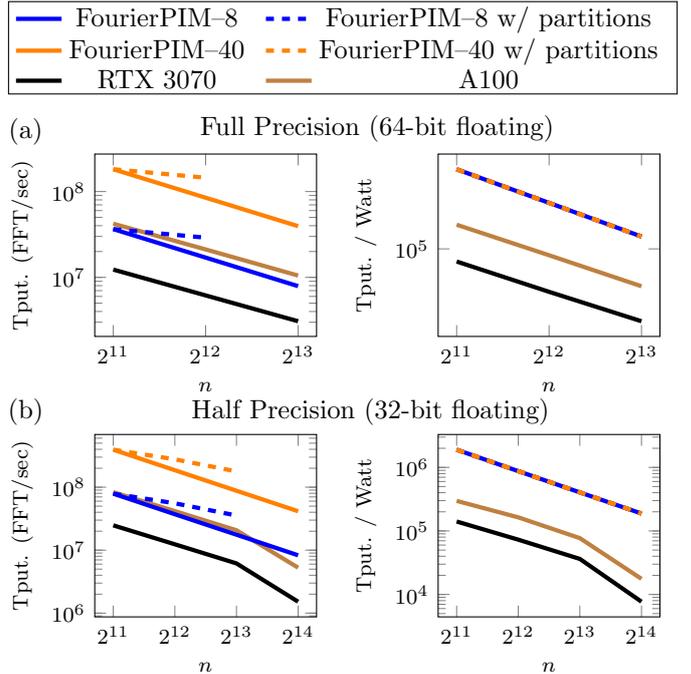

\begin{figure*}[t]
    \centering
  \begin{tikzpicture}
    \begin{groupplot}[
      group style={group size=4 by 2, horizontal sep=1.6cm,
      vertical sep=1.37cm},
      width=4.5cm, height=4cm
    ]
    \nextgroupplot[
        xmode=log,
        ymode=log,
        log basis x={2},
        xtick={2048, 4096, 8192},
        xlabel={\footnotesize $n$},
        ylabel={\footnotesize Tput. (Mult/sec)},
        every axis plot/.append style={ultra thick},
        every tick label/.append style={font=\footnotesize}
    ]
    \addplot[color=black]
    coordinates {(2048, 2742214.31)(4096, 1371266.59)};
    \addplot[color=brown]
    coordinates {(2048, 9327634.27)(4096, 4673064.9)};
    \addplot[color=blue]
    coordinates {(2048, 1.17E+07)(4096, 5.47E+06)};
    \addplot[color=blue, dashed]
    coordinates {(2048, 1.17E+07)(4096, 9.14E+06)};
    \addplot[color=orange]
    coordinates {(2048, 5.87E+07)(4096, 2.74E+07)};
    \addplot[color=orange, dashed]
    coordinates {(2048, 5.87E+07)(4096, 4.57E+07)};
    
    \coordinate (top5) at (rel axis cs:0,1);
    
    \nextgroupplot[
        xmode=log,
        ymode=log,
        log basis x={2},
        xtick={2048, 4096, 8192},
        xlabel={\footnotesize $n$},
        ylabel={\footnotesize Tput. / Watt},
        every axis plot/.append style={ultra thick},
        every tick label/.append style={font=\footnotesize}
    ]
    \addplot[color=black]
    coordinates {(2048, 17519.95)(4096, 8721.76)};
    \addplot[color=brown]
    coordinates {(2048, 42065.37)(4096, 20803.39)};
    \addplot[color=blue]
    coordinates {(2048, 2.01E+05)(4096, 9.24E+04)};
    \addplot[color=blue, dashed]
    coordinates {(2048, 2.01E+05)(4096, 9.24E+04)};
    \addplot[color=orange, dashed]
    coordinates {(2048, 2.01E+05)(4096, 9.24E+04)};
    \addplot[color=orange, dashed]
    coordinates {(2048, 2.01E+05)(4096, 9.24E+04)};
    
    \nextgroupplot[
        xmode=log,
        ymode=log,
        log basis x={2},
        xtick={2048, 4096, 8192},
        xlabel={\footnotesize $n$},
        ylabel={\footnotesize Tput. (Mult/sec)},
        every axis plot/.append style={ultra thick},
        every tick label/.append style={font=\footnotesize}
    ]
    \addplot[color=black]
    coordinates {(2048, 3499294.0)(4096, 1751763.4)(8192, 877056.9)};
    \addplot[color=brown]
    coordinates {(2048, 11884872.9)(4096, 5950771.8)(8192, 2968406.0)};
    \addplot[color=blue]
    coordinates {(2048, 1.62E+07)(4096, 7.60E+06)(8192, 3.57E+06)};
    \addplot[color=blue, dashed]
    coordinates {(2048, 1.62E+07)(4096, 1.21E+07)};
    \addplot[color=orange]
    coordinates {(2048, 8.12E+07)(4096, 3.80E+07)(8192, 1.78E+07)};
    \addplot[color=orange, dashed]
    coordinates {(2048, 8.12E+07)(4096, 6.05E+07)};
    
    \coordinate (top4) at (rel axis cs:0,1);
    
    \nextgroupplot[
        xmode=log,
        ymode=log,
        log basis x={2},
        xtick={2048, 4096, 8192},
        xlabel={\footnotesize $n$},
        ylabel={\footnotesize Tput. / Watt},
        every axis plot/.append style={ultra thick},
        every tick label/.append style={font=\footnotesize}
    ]
    \addplot[color=black]
    coordinates {(2048, 21807.6)(4096, 10860.2)(8192, 5695.5)};
    \addplot[color=brown]
    coordinates {(2048, 51803.0)(4096, 26073.1)(8192, 13132.2)};
    \addplot[color=blue]
    coordinates {(2048, 2.75E+05)(4096, 1.25E+05)(8192, 5.94E+04)};
    \addplot[color=blue, dashed]
    coordinates {(2048, 2.75E+05)(4096, 1.25E+05)(8192, 5.94E+04)};
    \addplot[color=orange, dashed]
    coordinates {(2048, 2.75E+05)(4096, 1.25E+05)(8192, 5.94E+04)};
    \addplot[color=orange, dashed]
    coordinates {(2048, 2.75E+05)(4096, 1.25E+05)(8192, 5.94E+04)};

    \nextgroupplot[
        xmode=log,
        ymode=log,
        log basis x={2},
        xtick={2048, 4096, 8192, 16384},
        xlabel={\footnotesize $n$},
        ylabel={\footnotesize Tput. (Mult/sec)},
        every axis plot/.append style={ultra thick},
        every tick label/.append style={font=\footnotesize}
    ]
    \addplot[color=black]
    coordinates {(2048, 5519324.3)(4096, 2760686.1)(8192, 1375143.6)};
    \addplot[color=brown]
    coordinates {(2048, 18552703.6)(4096, 9282465.4)(8192, 4620320.5)};
    \addplot[color=blue]
    coordinates {(2048, 2.57E+07)(4096, 1.21E+07)(8192, 5.74E+06)};
    \addplot[color=blue, dashed]
    coordinates {(2048, 2.57E+07)(4096, 1.79E+07)(8192, 1.14E+07)};
    \addplot[color=orange]
    coordinates {(2048, 1.29E+08)(4096, 6.07E+07)(8192, 2.87E+07)};
    \addplot[color=orange, dashed]
    coordinates {(2048, 1.29E+08)(4096, 8.96E+07)(8192, 5.69E+07)};
    
    \coordinate (top3) at (rel axis cs:0,1);
    
    \nextgroupplot[
        xmode=log,
        ymode=log,
        log basis x={2},
        xtick={2048, 4096, 8192, 16384},
        xlabel={\footnotesize $n$},
        ylabel={\footnotesize Tput. / Watt},
        every axis plot/.append style={ultra thick},
        every tick label/.append style={font=\footnotesize}
    ]
    \addplot[color=black]
    coordinates {(2048, 33042.5)(4096, 16984.1)(8192, 8435.3)};
    \addplot[color=brown]
    coordinates {(2048, 72849.8)(4096, 38896.6)(8192, 18557.2)};
    \addplot[color=blue]
    coordinates {(2048, 6.11E+05)(4096, 2.81E+05)(8192, 1.30E+05)};
    \addplot[color=blue, dashed]
    coordinates {(2048, 6.11E+05)(4096, 2.81E+05)(8192, 1.30E+05)};
    \addplot[color=orange, dashed]
    coordinates {(2048, 6.11E+05)(4096, 2.81E+05)(8192, 1.30E+05)};
    \addplot[color=orange, dashed]
    coordinates {(2048, 6.11E+05)(4096, 2.81E+05)(8192, 1.30E+05)};

    \nextgroupplot[
        xmode=log,
        ymode=log,
        log basis x={2},
        xtick={2048, 4096, 8192, 16384},
        xlabel={\footnotesize $n$},
        ylabel={\footnotesize Tput. (Mult/sec)},
        every axis plot/.append style={ultra thick},
        every tick label/.append style={font=\footnotesize}
    ]
    \addplot[color=black]
    coordinates {(2048, 7099474.3)(4096, 3548527.9)(8192, 1625429.5)(16384, 474287.0)};
    \addplot[color=brown]
    coordinates {(2048, 21677021.1)(4096, 11780061.5)(8192, 5031973.2)(16384, 1599154.6)};
    \addplot[color=blue]
    coordinates {(2048, 3.61E+07)(4096,1.71E+07)(8192, 8.10E+06)(16384, 3.84E+06)};
    \addplot[color=blue, dashed]
    coordinates {(2048, 3.61E+07)(4096,2.45E+07)(8192, 1.52E+07)};
    \addplot[color=orange]
    coordinates {(2048, 1.80E+08)(4096, 8.54E+07)(8192, 4.05E+07)(16384, 1.92E+07)};
    \addplot[color=orange, dashed]
    coordinates {(2048, 1.80E+08)(4096, 1.23E+08)(8192, 7.60E+07)};
    
    \coordinate (top2) at (rel axis cs:0,1);
    
    \nextgroupplot[
        xmode=log,
        ymode=log,
        log basis x={2},
        xtick={2048, 4096, 8192, 16384},
        xlabel={\footnotesize $n$},
        ylabel={\footnotesize Tput. / Watt},
        every axis plot/.append style={ultra thick},
        every tick label/.append style={font=\footnotesize}
    ]
    \addplot[color=black]
    coordinates {(2048, 39716.4)(4096, 20901.4)(8192, 9730.5)(16384, 2816.4)};
    \addplot[color=brown]
    coordinates {(2048, 82357.7)(4096, 46301.1)(8192, 20692.7)(16384, 6187.1)};
    \addplot[color=blue]
    coordinates {(2048, 8.40E+05)(4096, 3.89E+05)(8192, 1.81E+05)(16384, 8.49E+04)};
    \addplot[color=blue, dashed]
    coordinates {(2048, 8.40E+05)(4096, 3.89E+05)(8192, 1.81E+05)(16384, 8.49E+04)};
    \addplot[color=orange, dashed]
    coordinates {(2048, 8.40E+05)(4096, 3.89E+05)(8192, 1.81E+05)(16384, 8.49E+04)};
    \addplot[color=orange, dashed]
    coordinates {(2048, 8.40E+05)(4096, 3.89E+05)(8192, 1.81E+05)(16384, 8.49E+04)};

    \coordinate (bot) at (rel axis cs:1,0);
    
    \end{groupplot}
    
    \path (top|-current bounding box.north)--
          coordinate(legendpos)
          (bot|-current bounding box.north);
    \matrix[
        matrix of nodes,
        anchor=south,
        draw,
        inner sep=0.2em,
        draw
      ] at([xshift=-2ex,yshift=9ex]legendpos)
      {
    \ref{plots:PolyPIM8}& FourierPIM--8 &[5pt]
    \ref{plots:PolyPIM40}& FourierPIM--40 &[5pt]
    \ref{plots:RTX3070}& NVIDIA RTX 3070 &[5pt]\\
    \ref{plots:PolyPIM8P}& FourierPIM--8 w/ partitions &[5pt]
    \ref{plots:PolyPIM40P}& FourierPIM--40  w/ partitions&[5pt]
    \ref{plots:A100}& NVIDIA A100 &[5pt]\\};
    \node[font=\bfseries] at ([yshift=7ex, xshift=25ex]top5) {\Large Complex to Complex};
    \node[font=\bfseries] at ([yshift=7ex, xshift=23ex]top4) {\Large Real to Real};
    \node[] at ([yshift=2ex, xshift=25ex]top5) {Full Precision (64-bit floating)};
    \node[] at ([yshift=2ex, xshift=24ex]top4) {Full Precision (32-bit floating)};
    \node[] at ([yshift=2ex, xshift=24ex]top2) {Half Precision (16-bit floating)};
    \node[] at ([yshift=2ex, xshift=25ex]top3) {Half Precision (32-bit floating)};
    \node[] at ([yshift=2ex,xshift=-6ex]top4) {(c)};
    \node[] at ([yshift=2ex,xshift=-6ex]top5) {(a)};
    \node[] at ([yshift=2ex,xshift=-6ex]top2) {(d)};
    \node[] at ([yshift=2ex,xshift=-6ex]top3) {(b)};
  \end{tikzpicture}
  \caption{Comparison of (a, b) \emph{complex} and (c, d) \emph{real} polynomial multiplication to cuFFT on the NVIDIA RTX 3070 and A100. Notice that overlapping plots are shown in an alternating dashed format to highlight that both plots are present.}
  \label{fig:results:poly}
\end{figure*}

We evaluate the proposed algorithms compared to the NVIDIA GPU library cuFFT~\cite{cuFFT}. For the GPU baseline, we evaluate cuFFT on both an NVIDIA RTX 3070 and NVIDIA A100 (see Table~\ref{tab:params}) while measuring throughput using the \verb|cudaEvents| API and power using \verb|nvidia-smi|. The proposed algorithms are verified via a publicly-available cycle-accurate simulator\footnote{Available at https://github.com/oleitersdorf/FourierPIM.} that logically models a memristive crossbar array and performs the sequence of operations that correspond to the proposed algorithms.\footnote{As FourierPIM conforms to the abstract computational model (Figure~\ref{fig:PIM}), the validity of peripheral circuits follows from previous works~\cite{RACER, Nishil}.} The correctness is verified by inputting random data and comparing the output to the ground-truth outputs from the baseline implementations; performance is measured by counters in the simulator that record latency and energy. The algorithms are evaluated in a batched setup according to a digital memristive PIM architecture derived from Bitlet~\cite{Bitlet} and RACER~\cite{RACER}, as listed in Table~\ref{tab:params}, with both 8 GB and 40 GB variants (to match the RTX 3070 and A100 memory sizes, respectively). Further, we also evaluate FourierPIM with up to 4 partitions~\cite{PartitionPIM} to demonstrate the potential of partitions -- see Section~\ref{sec:proposed:2rbeta}.

We generated three benchmarks: (1) FFT, (2) polynomial multiplication, and (3) polynomial multiplication with real coefficients. The benchmarks consider a variety of dimensions $n \in \{2K, 4K, 8K, 16K\}$, corresponding to $2r, 2r\cdot 2,  2r\cdot 4$ and $2r \cdot 8$ algorithms, respectively\footnote{The dimensions are restricted in some cases (e.g., with partitions) due to the width of a single crossbar constraining intermediate memristor area.}, on both full-precision and half-precision floating-point numbers. The GPU batch size is chosen as the maximal value that fits in the GPU memory, and the batch size for FourierPIM is total the number of crossbars. We compare throughput (FFTs or polynomial multiplications per second) and throughput/Watt (energy).

Figure~\ref{fig:results:fft} shows the results of the FFT benchmark. For the full-precision algorithms, each complex number is a represented via $64$ bits (32-bit floating-point for each of $a$ and $b$). We notice that the throughput of both FourierPIM and cuFFT decrease approximately linearly in $n$, yet FourierPIM with partitions decreases logarithmically in $n$ (as the time complexity is $O(\log n)$). For example, for full-precision, we find a throughput improvement of up to $1.7\times$ using only two partitions and we expect these results to generalize to larger crossbars with more partitions.  This demonstrates the potential of memristive partitions to increase the throughput of PIM algorithms. Overall, for full-precision algorithms with partitions, we find an improvement of up to $5\times$ ($8\times$) in throughput (energy) compared to the RTX 3070, and an improvement of up to $7\times$ ($4\times$) in throughput (energy) compared to the A100. For half-precision numbers, we find improvements of $6\times$ (13$\times$) and $9\times$ ($6\times$) in throughput (energy) compared to the RTX 3070 and A100, respectively.\footnote{The decrease at $n=16K$ for cuFFT is part of a different linear trend that continues from $n=16K$ onwards since cuFFT performs differently on medium and large FFTs.} The increase in the ratios arises from GPUs only improving by $2\times$ (as they are memory bound and the data size decreases $2\times$) while PIM operates within the memory and thus reduces computation by a factor larger than $2\times$.

For the polynomial multiplication benchmarks, we would expect that the relative improvements remain roughly the same as FFT since both cuFFT and FourierPIM utilize the convolution theorem. Yet, we find that the relative improvement increases; for example, we find an improvement of up to $15\times$ and $10\times$ in throughput and energy for real half-precision polynomial multiplication compared to the A100, better than the FFT ratios of $9\times$ and $6\times$. This phenomenon is due to the fact that the element-wise multiplication performed by the GPU is also memory-bound, thereby contributing to the GPU latency and energy more than the contribution to PIM.

\section{Conclusion}
\label{sec:conclusion}

Fourier transforms are fundamental for a variety of applications ranging from signal processing to cryptography. Therefore, this paper seeks to accelerate the Fast Fourier Transform (FFT) by using PIM techniques to avoid the memory bottleneck. We begin by extending PIM arithmetic to complex numbers, continue by demonstrating a high-throughput in-memory butterfly operation, and then explore efficient mappings of the FFT to a crossbar array. We also extend the in-memory FFT to in-memory convolution and polynomial multiplication through the convolution theorem. The results are then evaluated through a publicly-available cycle-accurate simulator for the proposed algorithms that verifies correctness and measures performance, and the NVIDIA cuFFT library for a baseline on state-of-the-art GPUs. We find $\throughputOverGPU\times$ throughput and $\energyOverGPU\times$ energy improvement over state-of-the-art GPU. Multi-crossbar FFT will be investigated in the future.

\balance

\bibliographystyle{elsarticle-num}
\bibliography{refs}

\end{document}